\documentstyle[aps,epsfig,12pt,amssymb,amsmath]{revtex} 

\begin{document}

\title{Self-gravitating systems in a three-dimensional\\  expanding Universe }

\author{E.Aurell$^{1,2,3}$\thanks{e-mail: eaurell@sics.se} and D.~Fanelli$^2$\thanks{e-mail: fanelli@nada.kth.se}}

\maketitle 

\begin{center}
\begin{tabular}{ll}
$^1$ 	& SICS, Box 1263, SE-164 29 Kista, Sweden\\
$^2$ 	& Department of Numerical Analysis and Computer Science, KTH,\\ 
	& $\qquad$ S-100 44 Stockholm, Sweden\\
$^3$ 	& NORDITA, Blegdamsvej 17, DK-2100 Copenhagen, Denmark 
\end{tabular}
\end{center}

\begin{abstract} 
The non-linear evolution of stratified perturbations in a 
three-dimensional expanding Universe is considered. 
A general Lagrangian scheme is derived, and compared to
two previously introduced, approximate, models.
These models are simulated with heap-based event-driven
numerical procedure, that allows for the study of large systems, 
averaged over many realizations of random initial
conditions.
One of the approximate models is shown to be quantitatively similar to 
the adhesion model, as concerns mass aggregation. 
\end{abstract}


\section{Introduction}
\label{s:introduction}
Structure formation in the Universe is a rich and fascinating
problem, that touches many sides of Physics, from 
theories of the origin of primordial fluctuations, to detailed
calculations of galaxy dynamics in the present epoch, 
including radiation pressure from
stars and absorption in clouds, star birth and 
star mass loss. It is of great topical
interest today because the recent and still improving observational
data on the inhomogeneities in the cosmic microwave background
radiation.
\\\\
Structure formation is also a challenging problem of 
classical Physics, i.e. how a small perturbation
of a spatially almost uniform universe develops, first linearly and
then non-linearly, to the pronounced structures we see today.
The theory of the linear stage of this process was developed
by Lifshitz~\cite{Lifshitz} in the late 1940'ies, see \cite{Weinberg1972,ZeldovichNovikov},
while the nonlinear problem can either be treated as 
compressible hydrodynamics, or on the kinetic
level by N-body calculations or Vlasov equations.
In both cases, as always in non-linear hydrodynamics and gas dynamics
of large systems with many scales, approximations have to be made.
These typically amount to introducing turbulent viscosities
or diffusivities, either directly in modeling, or indirectly 
in the numerical scheme. Even so, large simulations of 
nonlinear gravitational instabilities are numerically
difficult problems, and a good deal of expertise and familiarity
with a scheme is required to evaluate the status and validity
of a result.
\\\\
For the reasons sketched above, there has since a long
time been an interest in
simplified models of structure formation. The very simplest starts from
the observation that in a one-dimensional setting, gravitational
attraction is a Lagrangian invariant between particle collisions.
This means that starting from an initially single-stream solution,
where velocity is a function of position, one can straight-forwardly
compute both the linear and the non-linear behavior up the moment of
caustic formation, and deduce e.g. the spectra
of density perturbations up to that time.
After caustic formation several
approaches are possible: either one may ignore the 
resulting change in the gravitational force (pancake, or Zeldovich model 
\cite{Zeldovich}),
or one may assume that its effects may be modeled by an
effective diffusive term (adhesion, or Gurbatov-Saichev-Shandarin model
\cite{GurbatovSaichevShandarin}).
Both these models can be solved in one, two or three dimensions,
and e.g. relations between statistics of initial conditions and
statistics of the solutions computed.
The last model, which gives Burgers' equation, leads already to interesting 
probabilistic and numerical problems, surveyed in \cite{Vergassola}.
\\\\
The present paper is the third in a series where we stay a little
closer to the original problem than working with Burgers' equation,
and study directly the collision-less self-gravitating media.
One motivation is to investigate if observed self-similar
behavior of the solutions of Burgers' equation also appears in
a self-gravitating system.
In~\cite{aurell}
we introduced a simplified model for investigating the structure formation in 
an expanding Universe. This discussion will be taken up here again,
in connection with the model of 
Rouet et al~\cite{MillerRouet,Rouet1990,Rouet1991} 
for the growth of perturbations in
a flat Universe.  In particular, we will derive a 
new model that is also suitable for numerical application 
({\it Quintic model}). In~\cite{al}, in collaboration with A.~Noullez, 
we presented an improved numerical algorithm, which takes ${\cal O}(\log N)$ 
operations per collision to simulate $N$ particles. This new tool, 
extensively adopted here, allows a significant enhancement of statistics
over our previous investigations.
\\\\
The paper is organized as follows. In section \ref{s:gr} we derive the 
Newtonian approximation, starting from General Relativity. In section
\ref{s:3d} we consider the special case of a stratified
perturbation in a $3D$ homogeneous expanding Universe. Then, in section 
\ref{s:FL}, we re-derive the approximate solution presented in 
~\cite{aurell}.  In section \ref{s:flat} we specialize to
a flat Universe and discuss the model 
of~\cite{MillerRouet,Rouet1990,Rouet1991},
and the new one presented here.
In section \ref{s:mass} we investigate properties of the mass distribution. 
Finally, in section \ref{s:fine} we sum up and discuss our results.

\section{From General Relativity to classical mechanics}
\label{s:gr}

In General Relativity, points in space and time manifold  
are labeled by four coordinates $x^{\mu}$. 
The Einstein conventions of covariant and contravariant
indices and summation over repeated indices are assumed.
The field $g_{\mu \nu}$ is (up to reparametrization) given by
the solutions of 
Einstein's field equations
\begin{equation}
\label{einstein}
R_{\mu \nu}-\frac{1}{2} g_{\mu \nu} R =8 \pi G T_{\mu \nu}
\end{equation}
where $R_{\mu \nu}$ is the curvature tensor,
$R = g^{\mu \nu}R_{\mu \nu}$ its trace, 
$T_{\mu \nu}$ the
energy-momentum tensor of the matter fields and
$G$ Newton's gravitational constant.
\\\\
For a perfect fluid moving with four-velocity
$U$ relative to a given coordinate system,
the energy momentum tensor is
\begin{equation}
\label{enmom}
T^{\mu \nu} = pg^{\mu \nu} + (p + \rho) U^\mu U^\nu
\end{equation}
where $p$ and $\rho$ are the pressure and the energy density.
Einsteins's equations simplify if the Universe is assumed
homogeneous and isotropic, and reduce then to Freidmann's
equation for a scale factor $a$
\begin{equation}
\label{eq:Friedmann}
{\dot a}^2 + k = \frac{8\pi G}{3} \rho a^2
\end{equation}
where $k$ is the sign of the curvature of three-dimensional 
slices ($-1$, $0$ or $1$).
A spatially isotropic and homogeneous metric $g$ is given 
by $a$, e.g. in the Robertson-Walker metric with
$g_{tt}$ equal to one, $g_{rr} = -a^2/(1-kr^2)$,
$g_{\theta \theta} = -a^2r^2$ and $g_{\phi \phi} = -a^2r^2\sin^2\theta$,
all off-diagonal elements zero.
In a curved Friedmann Universe ($k=1,-1$) $a(t)$ is its actual radius,
but in a flat Friedmann Universe ($k=0$) only the
ratios of $a$'s at different times have meaning.   
Equation~\eqref{eq:Friedmann} should be supplemented by the law of 
conservation of energy $(\rho a^3)' = -3pa^2$ and an equation of state, 
$p=p(\rho)$. 
\\\\
If the energy density is dominated by non-relativistic matter,
pressure can be neglected and $\rho(t) = \rho(t_0) (\frac{a(t)}{a(t_0)})^{-3}$.
If the Universe is flat ($k=0$) Friedmann's equation has the simple
solution $a(t) = a(t_0) (\frac{t}{t_0})^{\frac{2}{3}}$, where
the reference time $t_0$,
where $\rho(t_0)$ and  $a(t_0)$ are given,
satisfies $t_0=(6\pi G \rho(t_0))^{-\frac{1}{2}}$.
The present belief is that the Universe is actually flat or
close to flat. For the slightly more complicated expressions
that hold in an over-critical or an under-critical Universe, see e.g.
\cite{Weinberg1972}.
\\\\
Linear perturbations of the Einstein
equations (for $g$), around the Friedmann
solutions (where $g$ is given by $a$ and $k$), can be 
classified as scalar, vector and tensor, where
the scalar perturbations are coupled modes of density and potential
proper velocity, the vector modes are solenoidal proper velocity
fields, and the tensor perturbations are gravitational waves.
If we neglect gravitational waves, Friedmann's 
equations and the equations for the perturbations
can in fact be derived in a Newtonian setting, as explained
in \cite{Weinberg1972}, end of section~15.1,
which we will briefly recall here.
\\\\
Take a spherical part $S$ of the Universe, which
is assumed homogeneous and isotropic outside
$S$. The solutions of Einstein's equations in $S$ do
then not depend on the value of the density outside
$S$, just as there is no gravitational force acting
in Newtonian gravity on the inside of a spherical shell.
For small masses Einstein gravity is well approximated
by Newtonian gravity.
We first assume that mass density is constant in $S$,
equal to $\rho_b(t)$. If then $S$ is sufficiently small and
$\overrightarrow{r}$ is a coordinate system at rest with
respect to the Universe outside $S$, and with origin in the
center of $S$, a particle moves according to
\begin{equation}
\label{equation-of-motion}
m\frac{d^2\overrightarrow{r}}{dt^2} = -\frac{4\pi m G r^3}{3}\rho_b(t) \frac{1}{r^2}
\end{equation}
Introducing the proper coordinate $\overrightarrow{x}$
in $S$ such that
\begin{equation}
\label{scale1}
\overrightarrow{r} = \frac{a(t)}{a(t_0)} \overrightarrow{x} 
\end{equation}
we can rewrite \eqref{equation-of-motion} to
\begin{equation}
\label{classical-cosmological}
\frac{d^2 a}{dt^2} = - \frac{4\pi G}{3} \rho_b(t) a 
\end{equation}
Equation~\eqref{classical-cosmological} is the time-time component of the 
Einstein field equations (\ref{einstein}), assuming non-relativistic matter,
and also the time derivative of \eqref{eq:Friedmann}, assuming mass conservation.
\\\\
We now turn to perturbations, and to
simplify the notation we will write $a$ for $a(t)$, the
scale factor at time $t$, and $a_0$ and $\rho_0$ for $a(t_0)$ and 
$\rho_b(t_0)$, the scale factor and the background density at the
reference time $t_0$.
Let the density in $S$ at time $t$ fluctuate around the mean value
$\rho_b(t)$.
The Newtonian equations of motion for $N$
particles follow from a 
Lagrangian
\begin{equation}
\label{background_lagrangian}
{\mathcal{L}} = \sum_i \frac{1}{2} m_i \dot{\overrightarrow{r_i}}^2 - 
m_i \phi (\overrightarrow{r_i},t)
\end{equation}
where $\nabla_r^2\phi = 4\pi G\rho$.
By substituting the proper coordinate $\overrightarrow{x}$
and applying the canonical transformations~\cite{Peebles1980}: 

\begin{equation}
{\mathcal{L}} \rightarrow {\mathcal{L}} - \frac{d \Psi}{dt}~~~~~~,~~~~~~
\Psi =\sum_i \frac{1}{2} \frac{m a \dot{a}}{a_0^2} \overrightarrow{x_i}^2,
\end{equation}
we have
\begin{equation}
\label{lagrangian1}
{\mathcal{L}} =   \sum_i \frac{1}{2} m \frac{a^2}{a_0^2} \dot{\overrightarrow{x_i}}^2 - m \varphi(\overrightarrow{x_i})
\end{equation}
where $\varphi = \phi + \frac{1}{2} \frac{a \ddot{a}}{a_0^2} {\overrightarrow{x_i}}^2$.
The field equation for the new potential $\varphi$ reads:

\begin{equation}
\label{poisson1}
\nabla_x^2 \varphi  = 4 \pi G  \rho \frac{a^2}{a_0^2}+\frac{3 a \ddot{a}}{a_0^2},
\end{equation}
which, recalling (\ref{classical-cosmological}), can be written \cite{Peebles1980}:
\begin{equation}
\label{poisson2}
\nabla_x^2 \varphi  = 4 \pi G \left(\frac{a}{a_0}\right)^2 
\left( \rho( \overrightarrow{x},t)- \rho_b(t)\right).
\end{equation}
Equation~\eqref{poisson2} is the starting point of our work.
The source of $\varphi$ is not the density itself, but
the density perturbation  $\left(\rho-\rho_b(t)\right)$. This is as should
be: in the absence of 
density fluctuations each particle
moves according to the uniform expansion of the
Universe as a whole, but remains undisturbed in
the proper coordinate $\overrightarrow{x}$.
\\\\
The equation of motion associated to the Lagrangian (\ref{lagrangian1}) are
\begin{equation}
\label{equation-of-motion-Lagrangian}
\frac{d}{dt}\left(\frac{a^2}{a_0^2} \dot{\overrightarrow{x}}\right) + \overrightarrow{\nabla_x}\varphi(x)  = 0, 
\end{equation}
where, for simplicity, we neglect from here on the label $i$.
The peculiar velocity of a particle is not
$\dot{\overrightarrow{x}}$ 
but $\overrightarrow{v}=\frac{a}{a_0}\dot{\overrightarrow{x}}$. 
The equations of motion in Eulerian form, for the peculiar velocity
$\overrightarrow{v}$, are therefore
\begin{equation}
\label{euler}
\frac{d \overrightarrow{v}}{dt} + \frac{\dot{a}}{a} \overrightarrow{v}
= -\frac{a_0}{a} \overrightarrow{\nabla_{x}}\varphi.
\end{equation}
Equations~\eqref{euler} (or~\eqref{equation-of-motion-Lagrangian}),
~\eqref{poisson2}, and the equation of continuity define the
Newtonian model of structure formation we study. 
Closing this section, it is appropriate to summarize 
in what respects the model is inexact: i)~it ignores
gravitational waves, which are degrees of freedom of the
Einstein equations with no counter-parts in Newtonian theory;
ii)~it is limited to weak gravitational fields, in the sense
that Newtonian mechanics holds for the peculiar velocity;
iii)~it does not treat correctly the expansion of a local
patch with a density different from the average, since
density perturbations act as sources of {\it Newtonian}
(not {\it Einsteinian}) gravity.
Points~i) and~ii) are not serious for our purposes.
Point~iii) is actually not as problematic as it seems,
since the linear growth rates in the Newtonian model
agrees with the growth rates
in Lifshitz' full theory of linear perturbations to the
Friedmann solutions of the Einstein equations, so would
only be pertinent for the nonlinear growth of
perturbations of sufficiently long wave length.

\section{Stratified perturbations}
\label{s:3d}

We now consider the special case of a stratified perturbation: the velocity 
has one component only and varies with respect to this direction. Initial 
density also only varies in this direction. In the point particle picture, 
the density profile which goes into the Lagrangian 
\eqref{background_lagrangian} is 
\begin{equation}
\rho(x,t)=\sum_{x_j} m_j \left(\frac{a}{a_0}\right)^{-3} \delta(x-x_j).
\end{equation}
where $x$ is the comoving coordinate, in the direction of which
the density and velocity varies.
The Poisson equation leads to the following expression for the 
gravitational potential: 

\begin{equation}
\label{potform}
\varphi (x,t) = 2 \pi \left(\frac{a}{a_0}\right)^2 G\,  
 \int
dy|x-y| \left( \rho(y,t)- \rho_b(t) \right).
\end{equation}
The integral in \eqref{potform} should be taken 
over an interval from $-L/2$ to  $L/2$, $L$ eventually
taken to infinity, or to where the perturbation vanishes~\cite{aurell}.
Equation~\eqref{equation-of-motion-Lagrangian}
reads 
\begin{equation}
\label{euler1}
\frac{d^2 x}{dt^2} + 2 \frac{\dot{a}}{a} \frac{d x}{dt}
-4 \pi G  \rho_b(t) x = \left(\frac{a}{a_0}\right)^{-3} E_{grav}(x,t)
\end{equation}
where
\begin{equation}
\label{selfgrav}
E_{grav} (x,t) = - 2 \pi G  \sum_j m_j \hbox{sign} (x-x_j).
\end{equation}
The interesting thing is now that
by the equation of continuity
\begin{equation}
\rho_b(t) = \rho_0 \left(\frac{a(t)}{a(t_0)}\right)^{-3}
\end{equation}
the time-dependences of $4 \pi G  \rho_b(t)$
and $\left(\frac{a}{a_0}\right)^{-3} E_{grav}(x,t)$ are
actually the same.
As pointed out by Rouet et al~\cite{Rouet1990,Rouet1991},
all time-dependence can then be concentrated in the
term $2 \frac{\dot{a}}{a} \frac{d x}{dt}$ by a suitable
nonlinear transformation of the time variable.
The choice should be
\begin{equation}
\label{Rouet-time}
dt = \left(\frac{a}{a_0}\right)^{3/2} d \tau 
\end{equation}
where $\tau$ has dimension of time.
Equation (\ref{euler1}) is then transformed to
\begin{equation}
\label{euler3}
\frac{d^2 x}{d \tau^2} + \frac{\dot{a}\sqrt{a}}{2a_0^{3/2}} \frac{d x}{d \tau}
- 4 \pi G \rho_0  x = E_{grav} (x,\tau)
\end{equation}
The interest of this formulation is that, as in 
classical self-gravitating systems in one dimension,
$E_{grav}$ is a Lagrangian invariant,
proportional to the net mass difference to the right and to the 
left of a  given particle at a given time. 
Therefore, 
as far the particle do not experience any crossing, the equation of motion 
(\ref{euler3}) reduces to the compact form:

\begin{equation}
\label{compact}
\frac{d^2 x}{d \tau^2} + B(\tau) \frac{d x}{d \tau}
-  C x = const,
\end{equation}
where $B(t)= \frac{\dot{a}\sqrt{a}}{2a_0^{3/2}} $ and $C= 4 \pi G \rho_0$.
We have hence reduced the motion between collisions to a Ricatti equation.
When that can be integrated in an efficient manner, we can therefore
solve  general Newtonian model with a
fast event-driven scheme, as 
in~\cite{aurell,EldridgeFeix,MillerRouet,Rouet1990,Rouet1991}.
We will consider special cases when this can be done below.

\section{Approximate solution: FL model}
\label{s:FL}

\begin{figure}
\begin{center}
\psfig{figure=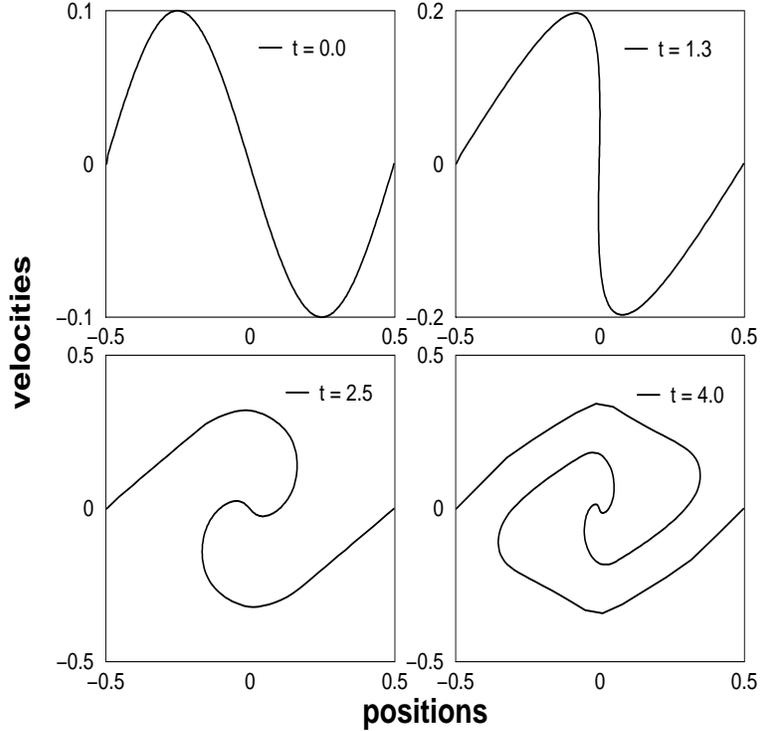,height=10truecm,width=10truecm}
\caption{\label{evolFL} Phase space portraits for the FL model with 
$N=200$. The plots refer to different $t$, expressed in unities of Jeans' 
time. The first plot shows the initial velocity profile. Positions and 
velocities are in arbitrary units.
}
\end{center}
\end{figure}

In our previous paper~\cite{aurell}
we introduced a model we here call Friction Less (FL) model.
That model basically cuts short the preceding discussion,
and starts directly from equation~\eqref{equation-of-motion-Lagrangian}
with the assumption that we consider only phenomena on time
scales much less than the age of the Universe.
This means that we can take $a$ constant, alternatively that
our time scales must be smaller than 
$a/\dot{a}$.
By a change of the spatial scale we can set $a$ to one,
and equation~\eqref{euler1} takes the much simpler form
\begin{equation}
\label{euler1-FL}
\frac{d^2 x}{dt^2} 
-4 \pi G  \rho_0 x =  E_{grav}(x,t)\qquad\hbox{FL model}
\end{equation}
Between collisions, \eqref{euler1-FL} has solutions
that are linear combination of two exponentials with rates $\pm \sqrt{4 \pi G \rho_0}$. This simple form allows the implementation of a fast and exact, 
up to round-off errors, integration scheme \cite{aurell,EldridgeFeix}: 
the evolution of the system is mapped from one crossing to the next and the 
times the events occur are analytically computed by solving a quadratic 
equation. In the present paper we use a version of the new heap-based  
algorithm introduced in \cite{al}. 
\\\\
In our numerical experiment we consider a periodic perturbation with size 
$2L$ with reflexion symmetry. Therefore we can simulate half of 
the symmetric system, confining $N$ particles in a box with reflecting 
edges in $-L/2$ and $L/2$. The mass of the particles is $m=N^{-1}$. 
We choose as unit of length the spatial 
interval in which the particles are initially distributed, and, thus, 
the initial density $\rho_0$ is equal to one. 
A natural choice of time scale is  
$\omega_{J0}^{-1}=\left(4 \pi G \rho_0\right)^{-1/2}=\sqrt{\frac{3}{2}} t_0$, 
where $\omega_{J0}$ is the Jeans frequency.
\\\\
In Fig. \ref{evolFL} we represent the evolution of the system  
in the phase space, starting with spatial uniform distribution. 
The initial velocity is a smooth function of position 
(first plots in Fig. \ref{evolFL}). As the time increases the multi-stream 
solution takes place. The spiral-like behavior develops and the thickness of 
the region where velocity is not single-valued grows quickly.

\subsection{Zeldovich approximation and FL model} 
\begin{figure}
\begin{center}
\psfig{figure=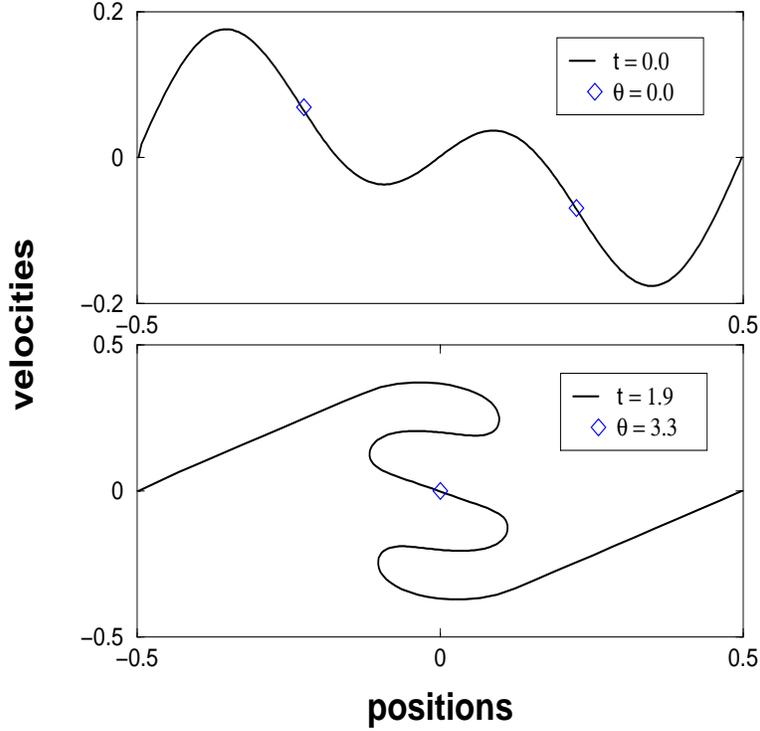,height=10truecm,width=10truecm}
\caption{\label{zeldo} The first plot shows the initial condition in the 
phase space. The two diamond are the macro-particles of mass $Nm/2$, initially
located in correspondence of the regions where the first singularities 
takes place. After $\theta=3.3 \omega_{J0}^{-1}$, the macro-particles have 
collided, and, as predicted, the multi-stream structures merge 
($t=\omega_{J0}^{-1} \sinh^{-1}(\omega_{J0} \theta)=1.9 \omega_{J0}^{-1}$).
Positions and velocities are in arbitrary units.}  
\end{center}
\end{figure}

As a test of the validity of the FL model we establish a connection 
with the Zeldovich approximation, closely parallel to the
similar relation for self-gravitating systems without 
expansion and Burgers' equation. For an extensive recent review 
on the subject the reader can refer to~\cite{Vergassola}. 
Consider the evolution of the system before the first crossing. 
According to eq. (\ref{euler1-FL}),  we have:

\begin{equation}
\label{compact1}
\frac{d^2x}{dt^2}= \omega_{J0}^2 (x-x_0),
\end{equation}
where $x_0=x(0)$. The solution of eq. (\ref{compact1}) can be written in 
the form:

\begin{equation}
x = x_0 + \frac{v_0}{\omega_{J0}} \sinh(\omega_J t)
\end{equation}
with $v_0=v(0)$. Following a suggestion by
S.N.~Gurbatov, we then introduce the new time defined as
$\theta=\frac{\sinh(\omega_{J0} t)}{\omega_{J0}}$ and the rescaled 
velocity $U = v/\dot{\theta} = v/\cosh(\omega_{0J} t)$
\cite{Gurbatov-private}. Thus, we get:

\begin{equation}
\label{burgers}
\begin{cases}
\begin{array}{l}
\displaystyle{x = x_0 + U(x_0) \theta}
\\
\\
\displaystyle{U(t,x_0)=U(x_0),}
\end{array}
\end{cases}
\end{equation}
which can be combined to give the Burgers equation:

\begin{equation}
\frac{\partial U}{\partial \theta} + U \frac{\partial U}{\partial x} = 0.
\end{equation}
Thus, the evolution of a particle can be thought as a free motion, 
by performing an appropriate rescaling of time and velocity,  
as far as no crossing takes place. 
As a less obvious consequence, the dynamics of 
the system can be described by
macro-particles, carrying a significant part of the whole mass of the system. 
These macro-particles are initially located at the points of future
caustic formation and evolve according to (\ref{burgers}), as far they 
do not cross each other. Thus, we have access to information on the times of  
merging of structures without investigating the details of their inner
dynamics.
\\\\
As an application consider the case of the smooth initial condition 
represented in Fig. \ref{zeldo}. The diamonds states for two macro-particles 
of mass $Nm/2$. After a time $\theta=3.3\omega_{J0}^{-1}$ the 
macro-particles have collided, and, as expected, the two multi-streamed 
structures have merged together. Time elapsed in original coordinate is 
$t=1.9\omega_{J0}^{-1}$.

\section{Models of a flat expanding universe}
\label{s:flat}

\begin{figure}
\begin{center}
\psfig{figure=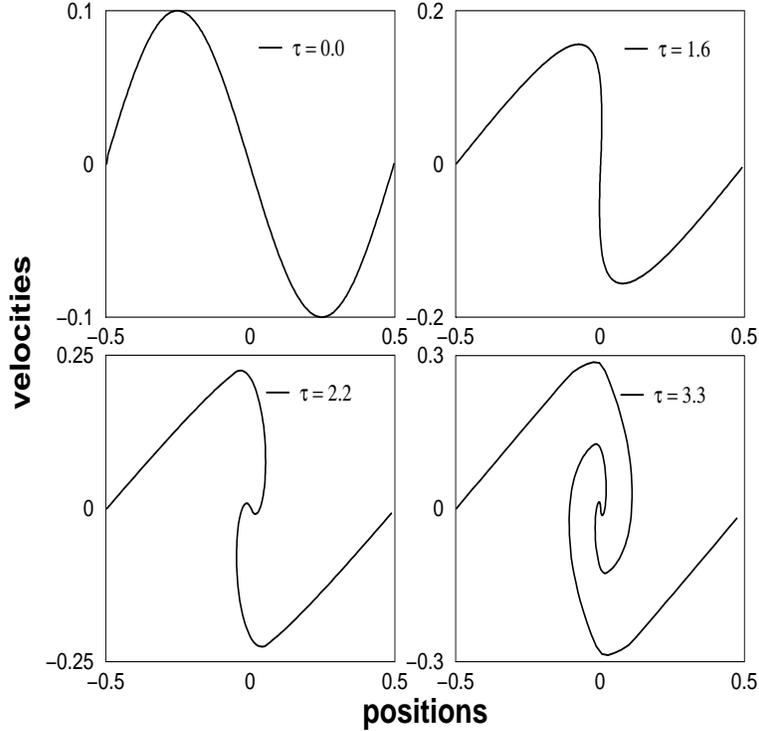,height=10truecm,width=10truecm}
\caption{\label{evolQ} Phase space portraits for Q model with $N=200$. 
The plots refer to different $\tau$,  specified in the figure legends in 
unities of Jeans' time. In terms of cosmological time, the times elapsed
after $t_0$ are respectively
$5.79 \omega_{J0}^{-1}$, $12.08 \omega_{J0}^{-1}$ and 
$46.47 \omega_{J0}^{-1}$.  
The first plot shows the initial velocity profile at $t_0$. Positions and 
velocities are in arbitrary units.
}
\end{center}
\end{figure}

In the matter-dominated era (pressure negligible) in a flat (critical) universe
($k=0$), the scale factor $a(t)$ grows with time 
as a power-law~\cite{Weinberg1972,ZeldovichNovikov}
\begin{equation}
a(t) = a_0 \left( \frac{t}{t_0} \right) ^{2/3}. 
\end{equation}
Here $t_0 = \left(6\pi G \rho_0\right)^{-1/2}$, as follows 
from equation (\ref{classical-cosmological}).
Therefore, as also pointed out by Rouet et 
al~\cite{Rouet1990,Rouet1991}, the sole remaining time-dependent 
term in \eqref{euler3} becomes time-independent. 
Let us note that the substitution  \eqref{Rouet-time} can be integrated to 
\begin{equation} 
\label{tau-t}
\tau = t_0  \log \frac{t}{t_0} \qquad
t = t_0 \exp\left(\frac{\tau}{t_0}\right)
\end{equation}
if we make the natural non-restrictive choice that $\tau(t_0)=0$.
Equation~\eqref{euler3} then takes the form:
\begin{equation}
\label{euler3-RF}
\frac{d^2 x}{d \tau^2} +  \frac{1}{3t_0} \frac{d x}{d \tau}
-  \frac{2}{3t_0^2} x =  E_{grav} (x,\tau) \qquad\hbox{Q model}. 
\end{equation}
We refer to  (\ref{euler3-RF}) as to the {\it Quintic (Q) model}.
Between collisions the 
right-hand side of (\ref{euler3-RF}) is constant and the homogeneous 
equation has solutions
\begin{equation}
\label{homogen}
x(\tau) = c_1 \exp(\frac{2(\tau-\tau_0)}{3t_0}) + c_2 \exp(-\frac{(\tau-\tau_0)}{t_0})
\end{equation}
where $c_1$ and $c_2$ are determined by $x(\tau_0)$ and $\dot{x}(\tau_0)$
in the transformed coordinate. 
The form of equation (\ref{homogen}) suggests introducing an auxiliary 
variable $z=\exp(\frac{(\tau-\tau_0)}{3t_0})$. The crossing times between 
consecutive particles are hence computed by solving numerically a  
{\it quintic equation} in the form:
\begin{equation}
b_1 z^5 - b_2 z^{3} + b_3 = 0,
\end{equation}
where the coefficients $b_1,b_2,b_3$ are fixed by the states of the particles 
at the time of the last crossing. Therefore, the event-driven scheme 
\cite{al} 
can be adopted to follow the dynamics of this system. 
In comparison, however, the FL model discussed above obviously leads to a 
quadratic equation, and the RF model presented below, which has been the 
benchmark in the literature, leads to a cubic equation. Both of these can be 
solved algebraically, while the quintic cannot.
The detail of the numerical implementation are given elsewhere
\cite{transp}. 
In the present numerical experiments we consider a system of $N$ 
particles, all of the same mass $m=N^{-1}$, confined in a finite box 
with reflective boundaries conditions. The unit of length is the box size 
and time is expressed in unit of the  inverse of the Jeans frequency  
$\omega_{J0}$.  
In Fig. \ref{evolQ} the evolution of the system  
in phase space is represented. The initial condition is the one we used for 
Fig. \ref{evolFL}. As is well-known, the time evolution leads to massive 
central core, and the system displays the expected spiral shape. 
Nevertheless, due to the action of the friction term, the particles remain 
well concentrated in the inner zone and the thickness of the multi-stream 
grows much slower than in the FL.

Let us note the fact, that if we re-express \eqref{homogen} in the original
time variable $t$ using (\ref{tau-t}), and assume small initial perturbations
of a uniform state, we then have, before any collisions,
\begin{equation}
\label{homogen1}
x(t) = \tilde{c}_1 \left( \frac{t}{t_0} \right)^{2/3} \,+\, 
\tilde{c}_2 \left( \frac{t}{t_0} \right)^{-1}.
\end{equation}
By the usual Eulerian-Lagrangian transformation,
this means that small density perturbations develop as
\begin{equation}
\label{homogen-density}
\delta\rho(x,t) = \rho_1(x,t_0) \left( \frac{t}{t_0} \right)^{2/3} \,+\, 
\rho_2(x,t_0) \left( \frac{t}{t_0} \right)^{-1}
\end{equation}
Similarly, small peculiar velocities transform according to
\begin{equation}
\label{homogen-velocity}
V(x,t) = v_1(x,t_0) \left( \frac{t}{t_0} \right)^{1/3} \,+ \,
v_2(x,t_0) \left( \frac{t}{t_0} \right)^{-4/3}
\end{equation}
In \eqref{homogen-density} and \eqref{homogen-velocity}
$(\rho_1,v_1)$ is the growing mode of coupled density
and velocity perturbations, while $(\rho_2,v_2)$ is
the decaying mode. 
It is easily checked that the growth rates in \eqref{homogen-density} and \eqref{homogen-velocity}
agree with the ones computed directly on the Eulerian side,
see~\cite{Weinberg1972}  or \cite{ZeldovichNovikov}.

\subsection{The Rouet-Feix Model (RF)}

\begin{figure}
\begin{center}
\psfig{figure=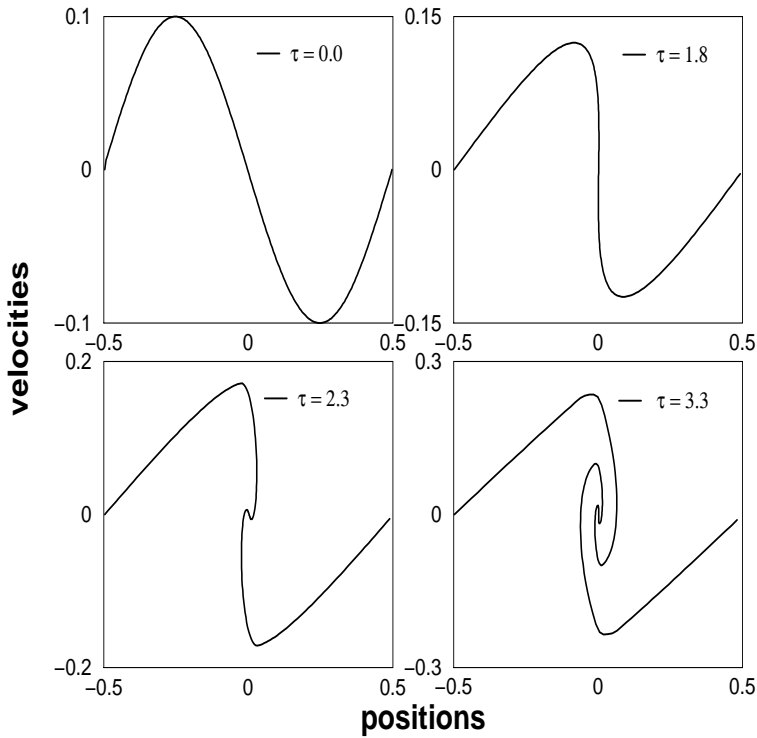,height=10truecm,width=10truecm}
\caption{\label{evolRF} Phase space portraits for RF model with $N=200$. 
The plots refer to different $\tau$,  specified in the figure legends in 
unities of Jeans' time. In terms of cosmological time, the times elapsed
after $t_0$ are respectively
$21.4632 \omega_{J0}^{-1}$, $61.9916 \omega_{J0}^{-1}$ and 
$517.14 \omega_{J0}^{-1}$.  
The first plot shows the initial velocity profile at $t_0$. Positions and 
velocities are in arbitrary units.
}
\end{center}
\end{figure}

A very interesting model was introduced about ten years ago by 
Rouet, Feix and collaborators~\cite{Rouet1990,Rouet1991},
see also the recent paper~\cite{MillerRouet}.
One way to understand this model, in the spirit of the original
derivation, is in the sense of a strictly Newtonian model of
an expanding universe.
Indeed, one may then specify an expansion
\begin{equation}
\label{RF-expansion}
a(t) = a_0 \left( \frac{t}{t_0^{1D}} \right) ^{2/3},
\end{equation}
without necessarily demanding, since one does not require compatibility
with general relativity, that the reference time $t_0^{1D}$
is connected to the density like in
$t_0=(6 \pi G \rho(t_0))^{-1/2}$.
We label the time  $t_0^{1D}$ since all change we actually have to do 
is to substitute \eqref{classical-cosmological} by the analogous 1D equation
\begin{equation}
\label{classical-cosmological1}
\frac{d^2 a}{dt^2} = - 4 \pi G \rho(t) a.
\end{equation}
From \eqref{RF-expansion} and \eqref{classical-cosmological1}
follow then
\begin{equation} 
\label{dens-t01d}
\rho=(18 \pi G t^2)^{-1} \qquad
t_0^{1D} = (18 \pi G \rho_0)^{-1/2}=\frac{\sqrt{2}}{3} \omega_{J0}^{-1}
\end{equation}
where $\omega_{J0}=(4\pi G \rho_0)^{1/2}$ 
\footnote{ Note that the numerical coefficient linking $t_0^{1D}$ to 
Jeans' time $\omega_{J0}^{-1}$ is equal to the inverse of the 
numerical constant $\alpha$ introduced by the authors in the derivation
in \cite{Rouet1991}.}. 
Equation (\ref{Rouet-time}) can be integrated to 
\begin{equation}
\label{tau-t1}
\tau = t_0^{1D}  \log \frac{t}{t_0^{1D}} = 
\frac{1}{\sqrt{3}} t_0 \log \left(\sqrt{3} \frac{t}{t_0} \right),
\end{equation}
which is linearly related to the $\tau$ given by
\eqref{tau-t}.
Between collisions equation~\eqref{euler3} is transformed into:
\begin{equation}
\label{rouet}
\frac{d^2 x}{d \tau^2} + \frac{\omega_{J0}}{\sqrt{2}} \frac{d x}{d \tau}
- \omega_{J0}^2 x = const. \qquad\hbox{RF model},
\end{equation}
which is the equation in \cite{Rouet1990,Rouet1991}.
The solution of the homogeneous equation takes the form:
\begin{equation}
\label{homogen-rouet}
x(\tau) = c_1 \exp(\frac{(\tau-\tau_0)\omega_{J0}}{\sqrt{2}}) + 
c_2 \exp(-\sqrt{2}(\tau-\tau_0)\omega_{J0}),
\end{equation}
which suggests the introduction of $z=\exp(\frac{(\tau-\tau_0)\omega_{J0}}{\sqrt{2}})$,
in terms of which one should solve the cubic equation
\begin{equation}
b_1 z^3 - b_2 z^2  + b_3 = 0
\end{equation}
to find the particle crossings. As in the previous case, the coefficients 
$b_1,b_2,b_3$ are fixed by the states of the particles at the time of the 
last crossing. 
We now transform back to time the coordinate $t$, using the second
equality of (\ref{tau-t1}), and find
\begin{equation}
\label{homogen2}
x(t) = \tilde{c}_1 \left( \frac{t}{t_0} \right)^{1/3} + 
\tilde{c}_2 \left( \frac{t}{t_0} \right)^{-2/3}.
\end{equation}
The exponents for the growing and decaying modes are different
than in the Lifshitz-Bonner theory.
Hence, the Rouet-Feix model does not predict correctly linear
growth of small perturbations, and can therefore not be taken
seriously as a quantitative model structure formation in a pressure-less
gas in an expanding universe, obeying general relativity.
\\\\
Nevertheless, the RF model has the important merit of producing a final 
tractable expression for computing algebraically the particle crossing times,
and has qualitatively all the right ingredients of the dynamics 
(both friction and background terms). 
It can therefore be simulated by a straight-forward generalization
of our heap-based event-driven scheme~\cite{al}.
For a more detailed discussion of the implementation, see
\cite{japan}.
\\\\
We note now, that at least in linear theory, there is another sense
one can give to the RF model. Namely, if $q$ is a wave-vector of a perturbation 
in a comoving frame ($\delta\rho \sim \rho_q \exp(i x q)$), and
we assume a non-negligible pressure $p\sim \rho^{\gamma}$ with
$\gamma=\frac{4}{3}$, then such perturbations grow as
\begin{equation}
\label{adiabatic}
\rho_q \sim t^{\alpha}\quad \alpha=-\frac{1}{6}\pm(\frac{25}{36}-\lambda^2)^{1/2}
\quad \lambda^2 = \frac{t^2 v_s^2 q^2}{a^2} 
\end{equation}
where $v_s$ is the sound speed, equal to $\sqrt{\frac{\gamma p}{\rho}}$.
These exponents can be matched with RF if $\lambda=2/3$. Hence, the RF model
describes the linear growth of perturbations of physical wave length 
$\frac{3}{2} t v_s$, which is $\sqrt{3/2}$ times larger than the instantaneous
Jeans length.
We note furthermore that the Jeans' length, $l_J = v_s/\sqrt{4\pi G\rho}$, scales
with time as $t^{2/3}$ under the assumptions above,
and that therefore the wave length of a perturbation remains in constant
proportion to the Jeans wave length. The RF model may hence well have some
validity also beyond linear theory. Let us add that the modifications of
the values of the exponents, of \eqref{homogen2} compared to
\eqref{homogen1}, are quite reasonable. As the length scale of a perturbation
approaches the Jeans length from above, the difference between expanding 
and decaying rates diminishes, to eventually give rise to purely oscillatory
behaviour below the Jeans' length.
\\\\
We now turn to simulations of the RF model.
We consider a system of $N$ particles, all of the same 
mass $m=N^{-1}$, confined in a finite box with reflective boundaries 
conditions.
The unit of length is the box size and time is expressed in unit of the 
inverse of the Jeans frequency  $\omega_{J0}$.  
In Fig. \ref{evolRF} we show the evolution of the system  
in the phase space, starting with the same initial condition we used for 
Fig. \ref{evolFL}. The result qualitatively agrees with the
plot of Fig. \ref{evolQ}. As for the Q model,  due to the presence 
of the friction, the system develops a multi-stream region, that is more 
narrow than the one displayed by the FL. Nevertheless it has to 
be noticed that the process of structure formation is considerably slower, 
for the RF, than for the Q model. Similar phase space portraits are 
displayed at equal rescaled $\tau$, but within the framework of the Q model, 
shorter cosmological times $t$ are required to attain such stages 
(see captions of Figs. \ref{evolQ}, \ref{evolRF}). 
In addition we observe that, for similar $\tau$,  the thickness of
the agglomeration is smaller in Fig \ref{evolRF} than in Fig. 
\ref{evolQ}. A simple explanation is, indeed, provided by looking to the 
relative importance of the friction and background terms,
respectively in Q and RF models. In the latter, in fact, the role of 
the friction is overestimated and, as a consequence, the process of particles
sticking is more pronounced. This observation suggests that, possibly, 
the adhesion model would result more similar to the approximate RF model, 
than to the exact Q model \cite{transp}.

\section{Mass octave function}
\label{s:mass}
\begin{figure}
\begin{center}
\psfig{figure=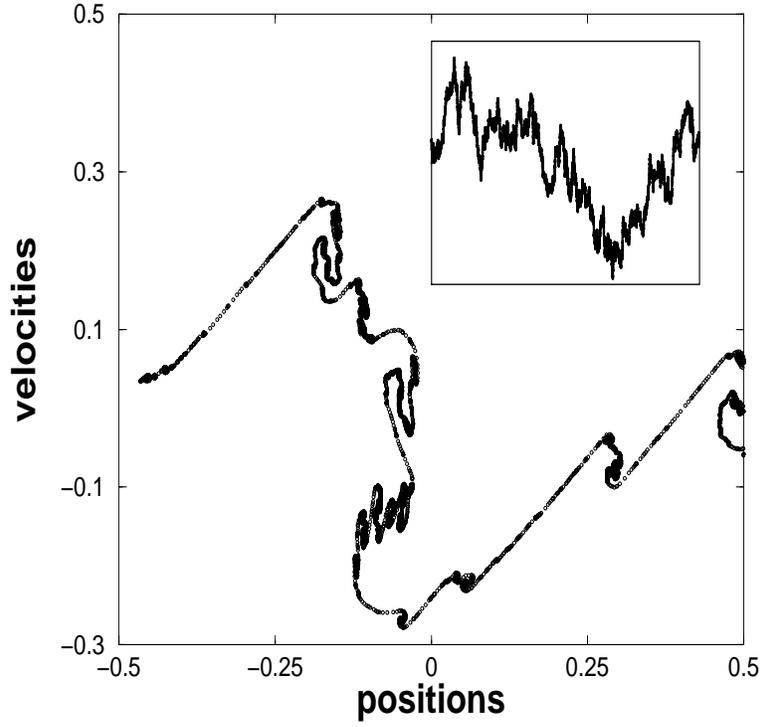,height=10truecm,width=10truecm}
\caption{\label{xvFL} Velocity field vs. position, for FL model, starting 
from a single-speed Brownian motion initial conditions (small inset). Here 
$N=16384$ and $t=4.0 \omega_{J0}^{-1}$. Reflecting boundaries are 
assumed. Positions and velocities are in arbitrary units.} 
\end{center}
\end{figure}

\begin{figure}
\begin{center}
\psfig{figure=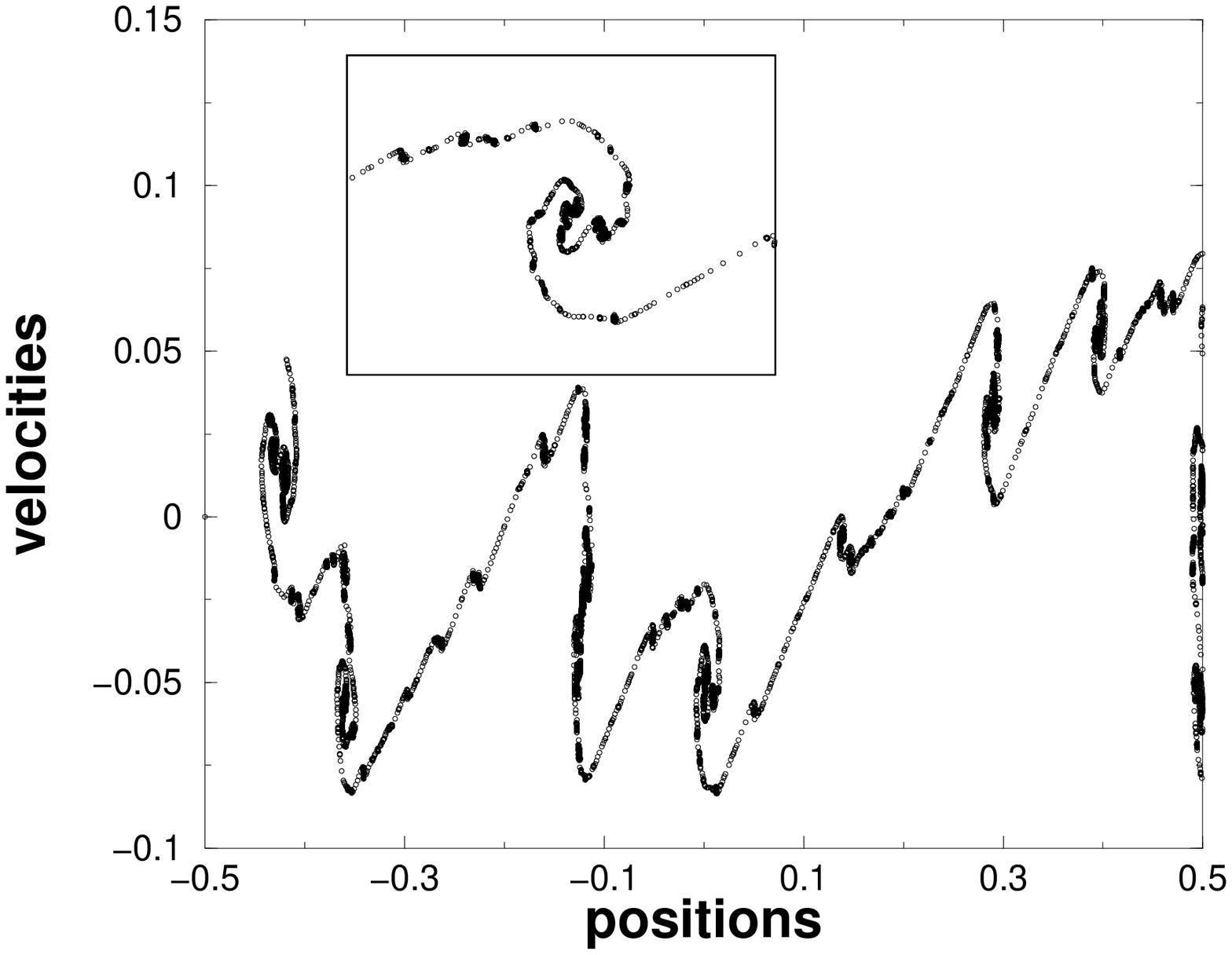,height=10truecm,width=10truecm}
\caption{\label{xvQ} Velocity field vs. position, for Q model,  starting 
from a single-speed Brownian motion initial conditions.
Here $N=16384$ and $\tau=4.0 \omega_{J0}^{-1}$ 
(or $t=109.53 \omega_{J0}^{-1}$).  Reflecting boundaries are 
assumed. The inset is a zoom of a massive agglomeration. Positions and 
velocities are in arbitrary units.}
\end{center}
\end{figure}

\begin{figure}
\begin{center}
\psfig{figure=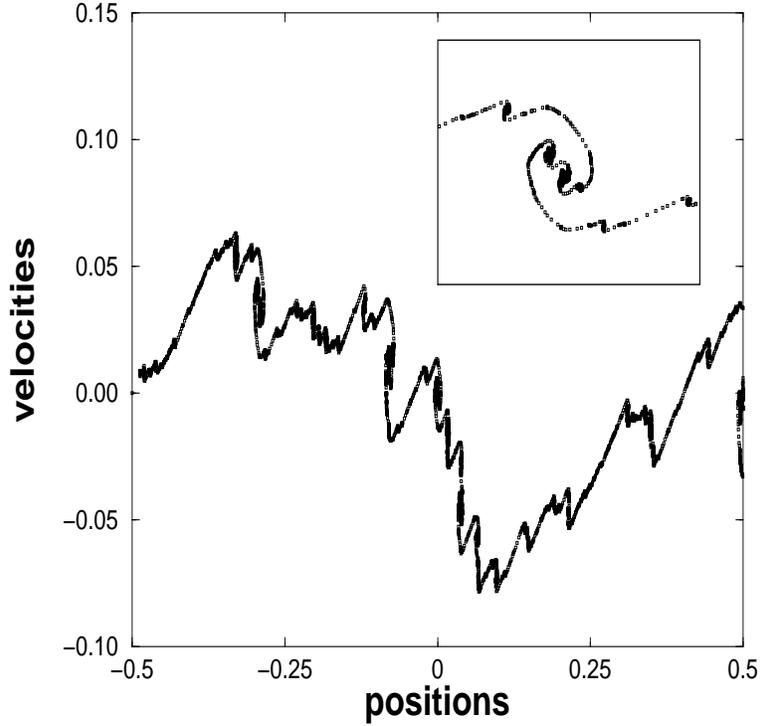,height=10truecm,width=10truecm}
\caption{\label{xvRF} Velocity field vs. position, for RF model, starting 
from a single-speed Brownian motion initial conditions.
Here $N=16384$ and $\tau=4.0 \omega_{J0}^{-1}$ 
(or $t=2.283 \times 10^3 \omega_{J0}^{-1}$).  Reflecting boundaries are 
assumed. The inset is a zoom of a massive agglomeration. Positions and 
velocities are in arbitrary units.}
\end{center}
\end{figure}

\begin{figure}
\begin{center}
\psfig{figure=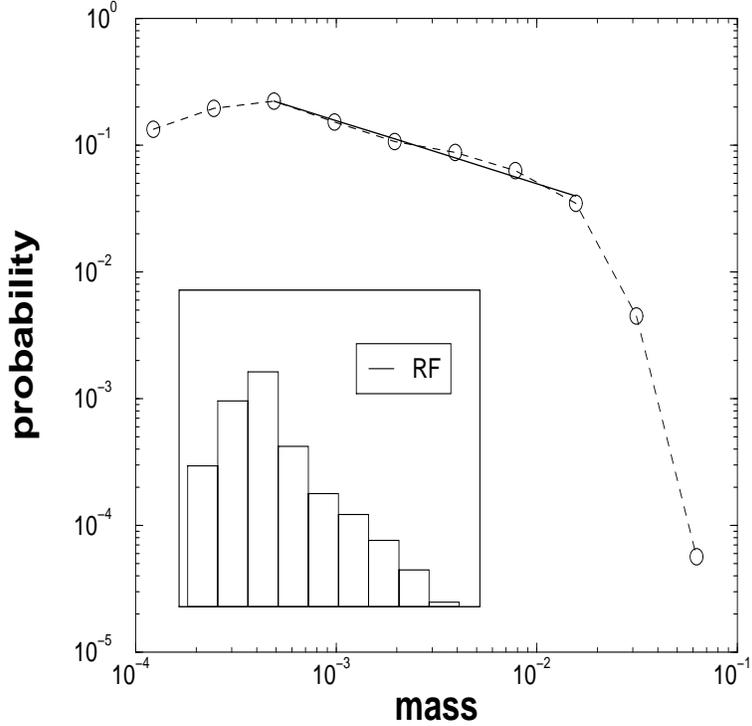,height=10truecm,width=10truecm}
\caption{\label{interm} Log-log normalized mass distribution in octaves (MOF) 
for Brownian initial velocities for RF (dashed line) at time  $\tau=3.3 
\omega_{J0}^{-1}$ (or $t=517.14 \omega_{J0}^{-1}$). 
The number of particles is $16384$, number of independent
realizations $1000$. The height of a column is the fraction of a total number 
of bins containing mass in the range $[m,2m]$. The bin size is $l=0.00195$, a
uniform density would hence correspond to a single column
at abscissa  $0.00195$ and ordinate $1.0$. An intermediate power-law regime 
is clearly displayed for RF model. The best numerical fit gives exponent 
$\alpha=-0.5$ (solid line). The small inset represents the 
MOFs in log-lin scale. Mass is in arbitrary units.
 }
\end{center}
\end{figure}

\begin{figure}
\begin{center}
\psfig{figure=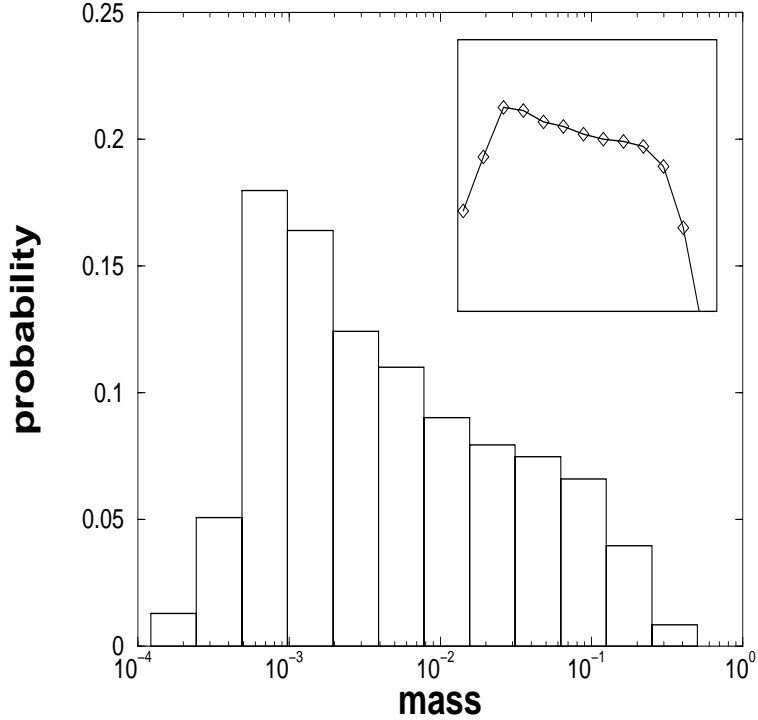,height=10truecm,width=10truecm}
\caption{\label{massRF} 
Normalized mass distribution in octaves (MOF) for 
Brownian initial velocities at $\tau=5.5 \omega_{J0}^{-1}$
(or $t=5.5007 \times 10^4 \omega_{J0}^{-1}$), for RF model.
The number of particles is $8192$, number of independent
realizations $1000$.
The bin size is $l=0.01562$. The small inset is a plot of the MOF in 
log-log scale. A numerical fit to a power law 
gives exponent $\alpha^{RF}=-0.25$. Mass is in arbitrary units.}
\end{center}
\end{figure}

\begin{figure}
\begin{center}
\psfig{figure=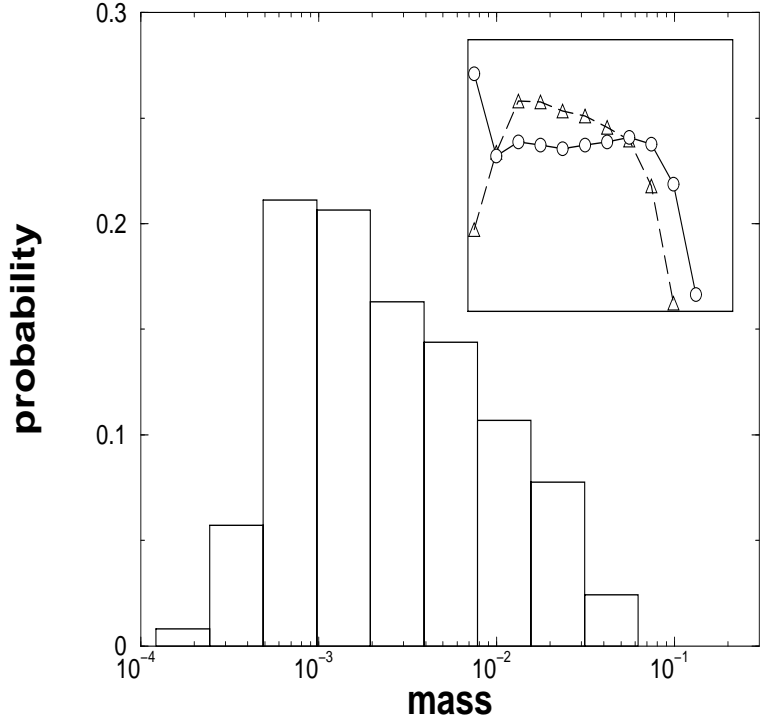,height=10truecm,width=10truecm}
\caption{\label{massQ} 
Normalized mass distribution in octaves (MOF) for 
Brownian initial velocities at $\tau=3.0 \omega_{J0}^{-1}$
(or $t= 32.18 \omega_{J0}^{-1}$), for Q model.
The number of particles is $8192$, number of independent
realizations $500$. The bin size is $l=0.003906$. 
The small inset represents the MOF in log-log scale for respectively
$t=3.0 \omega_{J0}^{-1}$ (dashed line and triangles) and 
$\tau=5.5 \omega_{J0}^{-1}$ ($t= 687.71 \omega_{J0}^{-1}$, solid line 
and circles). Bin size, number of 
particles and number of independent realizations as for the main figure.
An intermediate power-law regime, with positive exponent, develops as the 
evolution is pushed forward. The best numerical fit gives 
exponent $\alpha^{Q}=0.1$ at $\tau=5.5 \omega_{J0}^{-1}$ . 
Mass is in arbitrary units.}
\end{center}
\end{figure}

\begin{figure}
\begin{center}
\psfig{figure=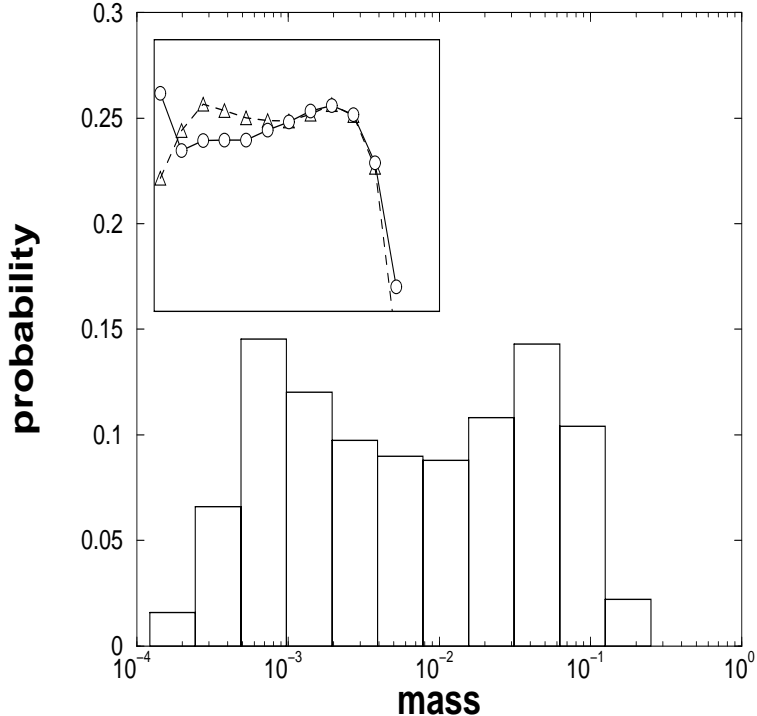,height=10truecm,width=10truecm}
\caption{\label{massFL} Normalized mass distribution in octaves (MOF) for 
Brownian initial velocities at $t=4.0 \omega_{J0}^{-1}$, for FL model.
Number of particles and number of realizations as in Fig. \ref{massRF}. 
The bin size is $l=0.01562$.  
The small inset represents the MOF in log-log scale for respectively
$t=4.0 \omega_{J0}^{-1}$ (dashed line and triangles) and 
$t=5.5 \omega_{J0}^{-1}$ (solid line and circles). Bin size, number of 
particles and number of independent realizations as for the main figure.
An intermediate power-law regime develops as the evolution is pushed forward.
The best numerical fit gives exponent $\alpha^{FL}=0.4$ at 
$t=5.5 \omega_{J0}^{-1}$ . Mass is in arbitrary units.} 
\end{center}
\end{figure}

In this section we study
the mass distributions for the 
FL, Q and RF models. Simulations are performed taking $N$ particles 
initially  uniformly distributed on a line of size $L$. Thus, the initial 
inter-particles distance is $\Delta x = L/N$. Velocities are generated as a 
Brownian random process. This is done in the Fourier space representation 
where:
\begin{equation}
v(x)=\Sigma_{k} v_k e^{ikx}
\label{randomfield}
\end{equation}
The sum over $k$ extends from $- \pi\,/\,\Delta x$ to 
$\pi \, /\,\Delta x$ in steps of $2 \,\pi \,/\,L$ and $v_{-k}=v_{k}^{*}$.
The Fourier components of positive $k$ are then chosen as a random Gaussian 
independent variables with variances:
\begin{equation}
<|v_k|^2>=\frac{\,\,k^{-2h-1}}{2\,\Delta x},
\label{variance}
\end{equation}
where $h$ is the Hurst exponent, the scaling exponent of the
second order structure functions.
The choice $h=0.5$ corresponds to 
standard Brownian motion.
The field generated by (\ref{randomfield}) and (\ref{variance})
will be periodic with period $L$. 
We scale $L$ to be one, and consider, as above, reflecting 
boundaries. 
\\\\
Fig. \ref{xvFL} shows a phase-space portrait for the FL model 
($t=4.0 \omega_J^{-1}$). Spiral structures are  displayed, as
well as stretched filaments connecting the dense agglomerations. 
Fig. \ref{xvQ} represents the phase-space for the Q model at
$\tau=4.0 \omega_J^{-1}$: the large-scale structures recall here even 
more the ones obtained in the framework of the adhesion model, even 
if it should be pointed out that, locally, multi-stream behaviour has 
in fact already started to develop (see inset in Fig. \ref{xvQ}). 
Substantially similar, from a merely qualitative point of view, is the 
phase space portrait for the RF model, at $\tau=4.0 \omega_J^{-1}$ 
(see Fig. \ref{xvRF}).
\\\\
To investigate quantitatively the particle 
distribution, we introduce the mass octave function (MOF), §
already discussed in \cite{aurell}.  MOF measures the probability to find a 
non-zero contribution to the mass density, as function of the mass itself, 
coarse-grained in octaves. In other words, $P(\Delta m)$ is the cumulative 
probability of finding $\Delta m$ in an octave interval $[m_k, m_{k+1}[$, 
where $m_k = 2^{-k}$~($k=0,1,$...). Practically, the MOF analysis reveals 
the degree of dishomogeneity of a probability distribution, since  in terms 
of the MOF a uniform distribution would correspond to a logarithmic histogram 
with only one non-zero entry. 
The main plot of Fig.~\ref{interm} show, in doubly
logarithmic scale,
the MOF computed for RF at an intermediate stage of 
evolution ($\tau=3.3 \omega_{J0}^{-1}$). 
In a finite range the mass octave function displays a power-law decay 
with exponent $-0.5$. This result is in agreement  with \cite{AFNB,She} 
where, in the framework of the adhesion model, the number density per unit 
length of shock locations holding mass $m$ is shown to be distributed as 
power-law $m^{-1-h}$. 
\\\\
Figs  ~\ref{massRF} show the MOF for a later time evolution of the RF model.
An approximate power law regime is found (see small inset), 
and a numerical fit gives exponent $\alpha^{RF}=-0.25$.
\\\\
The main plot of Fig  ~\ref{massQ} represents the MOF for 
the Q model at rescaled time $\tau=3.0 \omega_{J0}^{-1}$. 
There is no evidence of a transient power-law state, with slope $-0.5$, as 
shown for the RF. This observation again indicates that the adhesion picture 
is, in fact, closer to the RF model, than  to the exact Q approach. 
The small inset displays, in doubly logarithmic scale, the MOF for the Q 
model at $\tau=3.0 \omega_{J0}^{-1}$ and $\tau=5.5 \omega_{J0}^{-1}$. 
In the latter case an approximate power-law, over a finite mass support, is 
displayed and the best numerical fit gives exponent $\alpha^{Q}=0.1$.
\\\\
Figs.~\ref{massFL} shows on the other hand
the MOF for the FL model. At an intermediate stage,
($\tau=4.0 \omega_{J0}^{-1}$) the MOF is clearly not uniform,
but best described as a structure with two peaks.
At later times an approximate power-law regime is found,
with a positive exponent ($\alpha^{FL}=0.4$, at $\tau=5.5 \omega_{J0}^{-1}$). 
\\\\
We observe that the  exponent $\alpha^{Q}$ is practically equidistant 
from the values of $\alpha^{RF}$ and $\alpha^{FL}$. Nevertheless 
$\alpha^{Q}$ is positive, and therefore, concerning the mass aggregation, 
the exact Q dynamics is shown to be qualitatively more similar to the 
FL model, than to the RF. This is, indeed, a very surprising  
conclusion.

\section{Conclusions}
\label{s:fine}  
   
In this paper we have discussed the problem of structure formation   
in a three-dimensional expanding Universe, focusing on
stratified perturbations in a collision-less medium.
We derived a representation of the Lagrangian development
of such perturbations from the linear regime to large times, 
which forms the basis of an efficient numerical scheme ($Q$ model). 
In practice, this scheme requires solving automatically a large number of 
quintic equations, the coefficients of which are only known at run-time,
and therefore entails algorithmic problems that 
are discussed in \cite{transp}.
\\\\
We then re-derived as approximate solutions
the Friction Less Model (FL), recently introduced in \cite{aurell}, and  
the Rouet-Feix (RF) model, and compared them to the exact Q dynamics.
Quantitatively, the growth of small perturbations differs between RF and Q;
the RF model is in this sense more similar to growth of
finite wave length perturbations in a model with non-zero pressure.
Numerical simulations have been carried out for smooth and random initial 
perturbations, and phase space portraits, respectively for  FL, RF and Q, 
have been computed. The latters show compact collapsed structures, 
visually more like the shock waves (singular mass agglomerations) in the 
adhesion model (Burgers' equation).
\\\\
To make our analysis more quantitative, we studied the statistics of mass 
distributions from random initial conditions. For intermediate times, 
when the multi-stream behaviour is not well developed, the RF model 
agrees well with the scaling law for the frequency of agglomerations of 
given size containing mass $m$ in Burgers 
equation~\cite{Sinai,She,Vergassola}, i.e. $s(m)\sim m^{-1-h}$. 
This is not true for either the Q or the FL model. In particular, the FL 
seems to tend to a bifractal structure, with either very high or very 
low density agglomerations. 
\\\\
We finally studied the late time behaviour of the FL, Q and RF models.
All then display states which seem to have approximate power-law MOF. 
The actual states are however different,
with the most numerous mass agglomerations in FL (positive $\alpha^{FL}$) 
having about equal mass (they then also carry most of the mass), while the
most numerous mass agglomerations in the RF (negative $\alpha^{RF}$) are 
still small, and carry a small proportion of the mass. The exponent 
$\alpha^{Q}$, characteristic of the Q dynamics, is
practically equidistant from $\alpha^{RF}$ and  $\alpha^{FL}$. 
Nevertheless $\alpha^{Q}$ is positive and, therefore, surprisingly 
enough, we are led to conclude that, concerning the mass aggregation, 
the exact Q dynamics is qualitatively more similar to the FL model than to 
the RF model.
\\\\
Summing up, we have compared three models for studying the evolution of a 
stratified perturbation in an expanding universe. Only the Q model is
in quantitative agreement with the linear theory of growth 
of perturbations to the Friedmann solutions 
of the Einstein equations. Quantitatively, the approximate RF is shown to 
agree with the adhesion model, with respect to the mass aggregation.

\section*{Acknowledgments}
\label{s:acknowledgements}
We thank S.N.~Gurbatov, B.~Miller, P.~Muratore-Ginanneschi, A.~Noullez 
and J.L.~Rouet for discussions. This work was supported by the Swedish 
Research Council through grants NFR I~510-930 (E.A.) and NFR F~650-19981250 
(D.F).

\end{document}